\documentclass[a4paper, amsfonts, amssymb, amsmath, reprint, showkeys, nofootinbib, twoside]{revtex4-1}
\usepackage[english]{babel}
\usepackage[utf8]{inputenc}
\usepackage[colorinlistoftodos, color=green!40, prependcaption]{todonotes}
\usepackage{xcolor}

\usepackage[english]{babel}
\usepackage{xcolor}

\usepackage{amsthm}
\usepackage{mathtools}
\usepackage{physics}
\usepackage{graphicx}
\usepackage{adjustbox}
\usepackage{placeins}
\usepackage[T1]{fontenc}
\usepackage{lipsum}
\usepackage{csquotes}
\usepackage{soul}

\newcommand{\vk}[1]{{\color{purple} #1}}

\renewcommand{\vk}[1]{{\color{black} #1 }}

\newcommand*{\Ang}{\, \mbox{\normalfont\AA}} 

\newcommand{\avg}[1]{\ensuremath{\left<#1\right>}}

\newcommand{\udft}{\ensuremath{U}} %
\newcommand{\uml}{\ensuremath{U_{\text{MLP}}}} %
\newcommand{\umlhar}{\ensuremath{U_{\text{MLP}}^{\text{har}}}} %
\newcommand{\gdft}{\ensuremath{G}} %
\newcommand{\gml}{\ensuremath{G_{\text{MLP}}}} %
\newcommand{\aml}{\ensuremath{A_{\text{MLP}}}} %
\newcommand{\amlhar}{\ensuremath{A_{\text{MLP}}^{\text{har}}}} %
\usepackage[pdftex, pdftitle={Article}, pdfauthor={Author}]{hyperref} 
\begin{document}

\title{
A complete description of thermodynamic stabilities of molecular crystals
}

\author{Venkat Kapil}
\email[Correspondence email address: ]{vk380@cam.ac.uk}
\affiliation{Yusuf Hamied Department of Chemistry,  University of Cambridge,  Lensfield Road,  Cambridge,  CB2 1EW,UK}
\affiliation{Laboratory of Computational Science and Modeling, Institut des Mat\'eriaux, \'Ecole Polytechnique F\'ed\'erale de Lausanne, 1015 Lausanne, Switzerland}

\author{Edgar A Engel}
\email[Correspondence email address: ]{eae32@cam.ac.uk}
\affiliation{TCM Group, Cavendish Laboratory, University of Cambridge, J. J. Thomson Avenue, Cambridge CB3 0HE, United Kingdom}

\date{\today}

\begin{abstract}

Predictions of relative stabilities of (competing) molecular crystals are of great technological relevance, most notably for the pharmaceutical industry.
However, they present a long-standing challenge for modeling, as often minuscule free energy differences are sensitively affected by the description of electronic structure, the statistical mechanics of the nuclei and the cell, and thermal expansion. 
The importance of these effects has been individually established, but rigorous free energy calculations for \emph{general} molecular compounds, which simultaneously account for all effects, have \emph{hitherto} not been computationally viable. 
Here we present an 
efficient ``end to end'' framework that seamlessly combines state-of-the art 
electronic structure calculations, machine-learning potentials, and advanced free energy methods to calculate \textit{ab initio} Gibbs free energies for 
general organic molecular materials. 
The facile generation of machine-learning potentials for a diverse set of polymorphic compounds -- benzene, glycine, and succinic acid -- and predictions of thermodynamic stabilities in qualitative and quantitative agreement with experiments highlights that 
predictive thermodynamic studies of 
industrially-relevant molecular materials are no longer a daunting task. 

\end{abstract}


\maketitle


\section*{Introduction}

Molecular crystals are ubiquitous in the pharmaceutical industry~\cite{datta_crystal_2004} and show great promise for applications in organic photovoltaics~\cite{forrest_path_2004}, 
gas adsorption~\cite{tozawa_porous_2009}, and the food, pesticide and fertilizer industries~\cite{honer_mechanosynthesis_2017}.
Their tendency to exhibit polymorphism, i.e. to exist in multiple crystal structures, on one hand provides a mechanism to tune properties by controlling crystal structure~\cite{gentili_polymorphism_2019}, and on the other hand introduces the challenge of synthesising and stabilizing crystal structures with desired properties~\cite{chistyakov_polymorphism_2020}.
While thermodynamic stability at the temperature and pressure of interest is sufficient (although not necessary\footnote{kinetics may protect thermodynamically metastable structures from decaying almost indefinitely}) to ensure long-term stability, simply understanding thermodynamic stability already poses a formidable challenge.
This is particularly true for pharmaceuticals, where free energy differences between drug polymorphs are often smaller than 1\,kJ/mol~\cite{cruz-cabeza_facts_2015}, leading to the risk of the drug transforming 
into a less soluble and consequently less effective form
during manufacturing, storage or shelf-life~\cite{chemburkar_dealing_2000, chaudhuri_crystallisation_2008}. 
Indeed, the problem of late appearing drug polymorphs is widespread~\cite{rietveld_rotigotine_2015, bucar_disappearing_2015}. \\

The pharmaceutical industry therefore spends considerable resources on high-throughput crystallization experiments to screen for polymorphs~\cite{morissette_high-throughput_2004}, into which the target structure may decay. 
However, crystallization experiments do not probe thermodynamic stability, and conclusive studies of the impact of temperature changes after crystallisation on the stability of polymorphs (i.e. their monotropic or enantiotropic nature~\cite{lee2014}) are often prevented by limited sample quantities. 
Hence the appeal of theoretical crystal structure prediction (CSP)~\cite{price_predicting_2014} based on the thermodynamic stability,
which promises to complement crystallization experiments~\cite{nyman_crystal_2018} by exhaustively searching for competing polymorphs. \\

Despite the demonstrable value of CSP for many classes of materials~\cite{pickard2007, oganov2009, pickard2015, zhuang, marzari2018, conduit2017}, and the continuing progress evidenced by a series of blind tests~\cite{nyman_crystal_2018}, the success of CSP for molecular crystals has been limited by the inability to routinely predict the relative stability of competing candidate structures~\cite{reilly_report_2016}. 
This is largely because the methods used for stability rankings typically ignore or approximate the subtle interplay of several effects, such as intricate inter-molecular interactions~\cite{marom_many-body_2013}, the (quantum) statistical mechanics of the nuclei~\cite{rossi_anharmonic_2016} and the unit cell~\cite{schneider_exploring_2016}, and thermal expansion~\cite{ko_thermal_2018}, thereby incurring errors larger than the free energy differences of interest. 
The importance of each of these effects has been demonstrated in isolation, but predictive stability rankings must also comprehensively account for their interplay. \\

Recent implementations of advanced path-integral (PI) approaches~\cite{markland_nuclear_2018, kapil_i-pi_2019} allow exactly accounting for the quantum statistical mechanics of the nuclei and the unit cell~\cite{kapil_assessment_2019, kapil_modeling_2019} for arbitrary potential energy surfaces (PESs). 
At the same time, modern machine learning potentials (MLPs)~\cite{deringer2019} 
permit accurately reproducing \emph{ab initio} PESs and dramatically reduce the cost of performing simulations approaching \emph{ab initio} accuracy~\cite{lan_simulating_2021}. 
Despite these advances, calculations of rigorous thermodynamic stabilities for general molecular materials have been complicated by the absence of a integrated framework, which facilitates both the rapid development of MLPs and free energy calculations including all physically relevant effects, while ensuring universal applicability to diverse systems. \\

\begin{figure*}
    \centering
    \includegraphics[width=\textwidth]{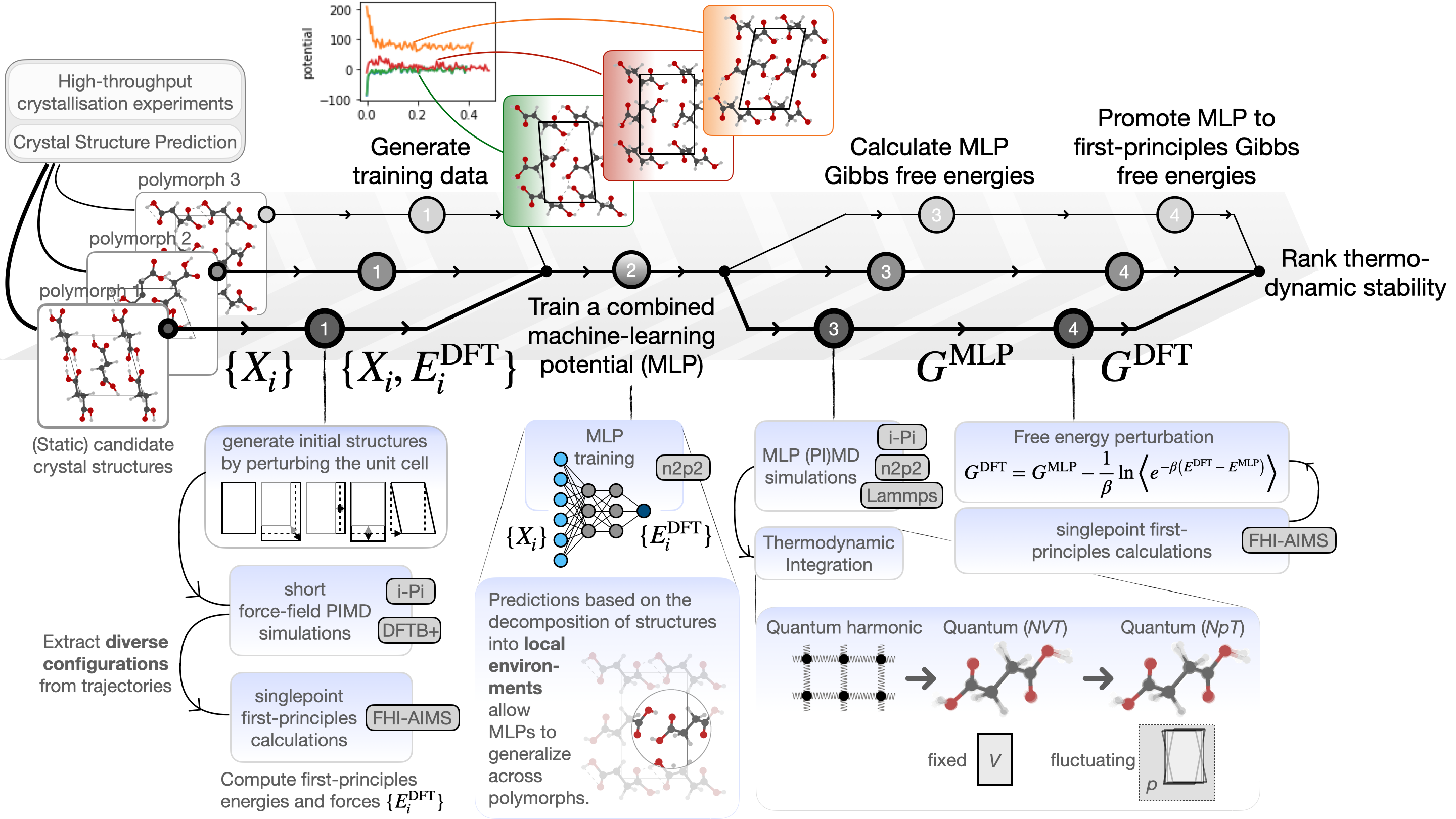}
    \caption{Schematic representation of the workflow for computing \textit{ab initio}, quantum anharmonic Gibbs free energies for candidate crystal structures. 
    The upper half of the figure shows the main steps: (1) generating \textit{ab initio} reference data on which to (2) train a combined MLP, which can then be used to (3) compute MLP Gibbs free energies, which one can finally (4) promote to \textit{ab initio} Gibbs free energies.
    The lower half (shaded in blue) details the key aspects of how each of these steps is performed in practice.}
    \label{fig:workflow}
\end{figure*}

\begin{figure}[t]
    \centering
    \includegraphics[width=0.35\textwidth]{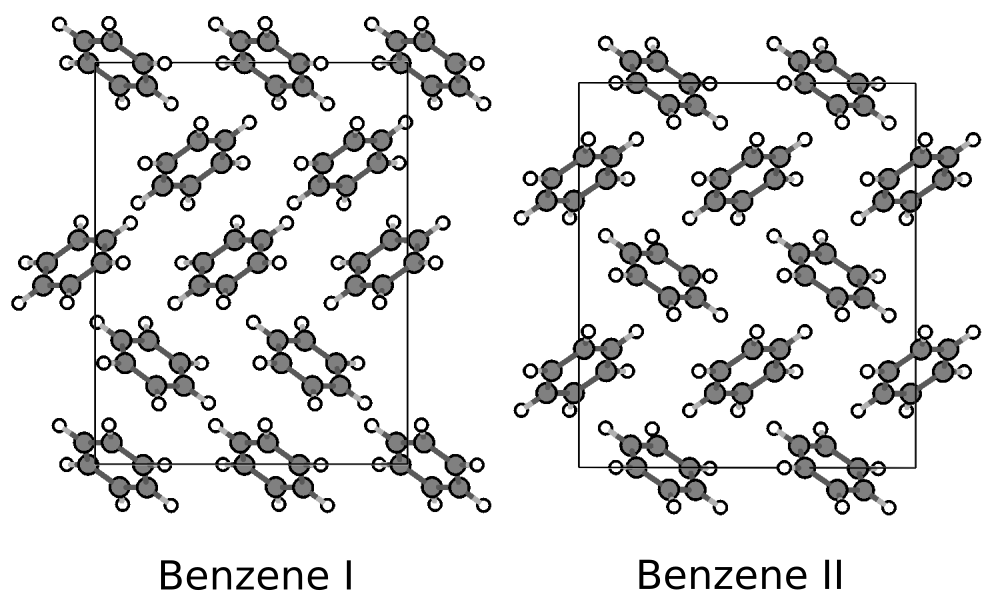}\\
    \vspace{5pt}
    \includegraphics[width=0.45\textwidth]{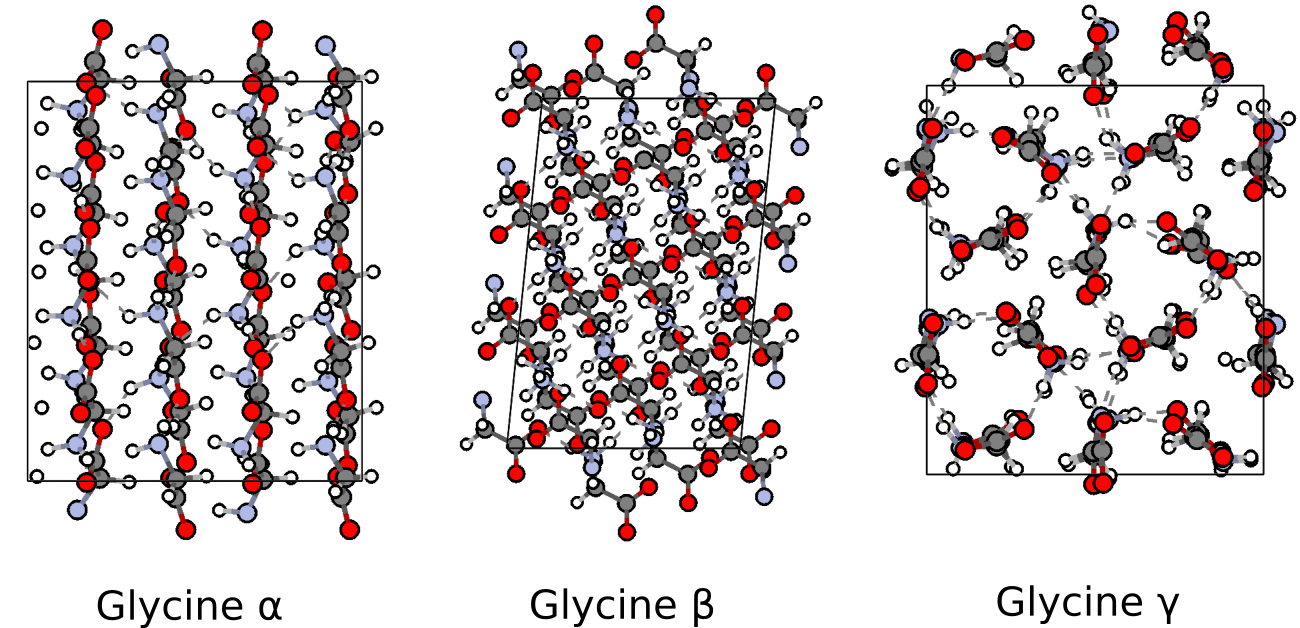}\\
    \vspace{5pt}
    \includegraphics[width=0.35\textwidth]{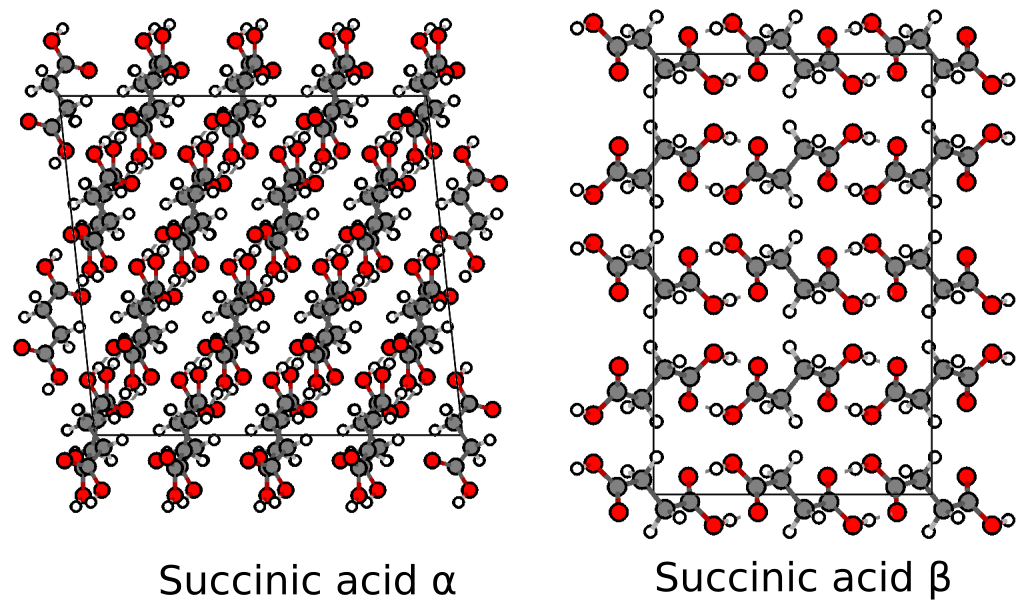}\\
    \caption{Structures of forms I and II of benzene containing 16 molecules, forms $\alpha$, $\beta$, and $\gamma$ of glycine containing 24 molecules, and forms $\alpha$ and $\beta$ of succinic acid containing 24 molecules. Hydrogen, carbon, nitrogen, and oxygen atoms are shown in white, gray, blue, and red, respectively.}
    \label{fig:polymorphs}
\end{figure}

In this work we present an efficient framework for ranking candidate structures of arbitrary compounds using rigorous \emph{ab initio} Gibbs free energy calculations, based on the streamlined development of MLPs and their integration with PI methods.
Our approach builds upon our previous work on combining PI approaches with MLPs for ice polymorphs~\cite{kapil_assessment_2019, cheng_ab_2019}, but greatly enhances its accuracy, efficiency, and robustness for out-of-the-box applications to general compounds.
In particular, we simplify the development of MLPs using a straightforward and inexpensive protocol for compiling \textit{ab initio} reference data, which is designed to work for general organic compounds and accounts for (the often-neglected) cell flexibility and quantum nuclear motion. 
Additionally, robust data-driven techniques minimize the human effort involved in training the MLPs. 
In contrast to previous CSP ranking methods that use MLPs~\cite{wengert_data-efficient_2021, mcdonagh_machine-learned_2019}, we exactly account for the \emph{quantum} statistical mechanics of the nuclei and the cell, and only use MLPs as a stepping stone for computing \emph{ab initio} Gibbs free energies, eliminating all dependence on the MLPs and their limitations.
The reliability and general applicability of our approach is showcased by the rapid development of MLPs 
and correct stability predictions for crystal polymorphs of three prototypical compounds: benzene, glycine, and succinic acid. These bear the hallmarks of more complex bio-molecular systems -- molecular flexibility, competing polymorphs, and inter-molecular interactions ranging from weak dispersive, to hydrogen bonded and ionic. Importantly, the relative stability of their polymorphs is well-established~\cite{katrusiak2010, leviel1981, perlovich_polymorphism_2001}. 
We further assess the temperature and pressure dependence of relative stabilities based on gradients of Gibbs free energies, which correspond to indicators widely used by experimentalists to predict the monotropic or enantiotropic nature of the polymorphs. \\
Our work complements state-of-the-art CSP methods, which efficiently survey structural space to extract small sets of promising candidate structures using \textit{ab initio} calculations and/or MLPs~\cite{wengert_data-efficient_2021, mcdonagh_machine-learned_2019}, but struggle to reliably resolve subtle differences in stability among them~\cite{reilly_report_2016}. 
Combining rigorous free energy calculations, as demonstrated here, with structure searching and inexpensive CSP ranking methods, constitutes an avenue to predictive CSP for complex molecular crystals of industrial importance. \\

\section*{Computational framework and systems}

To predict rigorous relative stabilities, we combine PI thermodynamic integration~\cite{kapil_assessment_2019} (referred to as QTI) in the constant pressure ensemble (thereby accounting for anharmonic quantum nuclear motion and the fluctuations and thermal expansion of the cell) with density-functional-theory (DFT) calculations with the hybrid PBE0 functional~\cite{perdew1996pbe0, adamo1999pbe0} and the many-body dispersion correction of Tkatchenko \emph{et al.}~\cite{tkatchenko2012mbd, ambrosetti2014mbd} (referred to as PBE0-MBD). 
PBE0-MBD provides an accurate description of inter-molecular interactions, as benchmarked using experimental and CCSD(T) lattice energies for various molecular crystals, including form I of benzene and $\alpha$-glycine~\cite{reilly2013, beran2016}. 
Since direct calculation of Gibbs free energies using \textit{ab initio} QTI is prevented by the cost of the required energy and force evaluations~\cite{kapil_assessment_2019}, \textit{ab initio} Gibbs free energies are calculated in a four-step process, as depicted schematically in Fig.~\ref{fig:workflow} and detailed further in the supporting information (SI). \\

First, we use a simple strategy to generate a minimal but exhaustive set of \emph{unit cell} ``training configurations'', for which we then perform PBE0-MBD calculations:
we perform PI simulations based on density-function tight binding (DFTB)~\cite{elstner1998} theory for unit cells with perturbed cell parameters. This allow us to gather a large number of configurations, which incorporate quantum nuclear fluctuations and cell flexibility, and from which we can distill  
the most distinct ones using a data-driven approach~\cite{imbalzano_automatic_2018}.
This strategy leverages the low cost of DFTB and its qualitative accuracy for diverse molecular crystals~\cite{brandenburg_accurate_2014} to avoid the bottleneck that is PBE0-MBD-based configurational sampling.  
Thanks to the versatility of DFTB, it can be used to generate robust training data for almost any compound of interest. 
The subsequent training of the MLPs hinges on identifying   
the most important ``features'' of the configurations, fed to the MLPs as input. 
These features are usually abstract functions quantifying the local density of atoms, and require the careful tuning of multiple parameters~\cite{imbalzano_automatic_2018}.
Here, we render training MLPs for general compounds accessible to non-experts by automating this procedure  using a ``size extensive'' data-driven approach, which avoids the manual selection of features based on ``prior experience''.
Combining these first two steps 
with a ``tried and tested'' neural network architecture~\cite{behl-parr07prl, behl11pccp, kapil_high_2016} greatly simplifies and speeds up the generation of MLPs, while remaining agnostic to the system of study. \\

In a third step, we exploit the orders-of-magnitude lower cost of the resultant MLPs compared to the \emph{ab initio} reference method, to compute Gibbs free energies for much larger simulation supercells using QTI~\cite{kapil_assessment_2019}. 
We account for anisotropic fluctuations of the simulation cell, which are important for flexible functional materials~\cite{lee2021}, and 
directly calculate the free energy difference between the harmonic reference systems and the physical, anharmonic system at the PI level, which substantially reduces the complexity and cost compared to the multi-step integration performed in Ref.~\cite{cheng_ab_2019}. 
We note that the affordability of MLP free energies comes at the price of residual errors with respect to the \emph{ab initio} reference values due to the imperfect reproduction of the reference PES.
These may arise from the short ranged nature of the MLPs~\cite{grisafi_incorporating_2019}, information lost during the ``featurization'' of the configurations~\cite{pozdnyakov_incompleteness_2020}, or from insufficient training data.
The typical errors in MLP predictions of configurational energies (Table~\ref{tab:errors}) are small but comparable to the subtle free energies differences between polymorphs. 
Therefore, in a fourth and final step, we eliminate the associated errors to obtain true \emph{ab initio} Gibbs free energies by computing the difference between the MLP and PBE0-MBD free energies using FEP~\cite{cheng_ab_2019}.
All calculations and simulations are performed using readily-available and well-documented software, and Jupyter notebooks for analysis are provided as supporting material. \\

As an expos{\'e} of the universal applicability of this scheme, we predict the relative stabilities of a set of prototypical systems, whose small number and size belies how representative they are of general organic molecular crystals: benzene is the archetypal rigid, van-der-Waal's bonded molecular crystal, while succinic acid represents general hydrogen-bonded systems, and glycine prototypes flexible zwitter-ionic systems. 
This small, ``irreducible'' set of prototypical systems not only covers the three different types of bonding, but also the chemical space that includes pharmaceuticals such as aspirin and paracetamol. 
Moreover, molecular flexibility and the large amplitude curvilinear motion of the amide group in glycine trigger the same pathologies of approximate free energy methods as more complex systems exhibiting free rotation of molecular units~\cite{rossi_anharmonic_2016, kapil_assessment_2019}, and serve as a stringent test for stability predictions.

For each compound we compute the free energy differences between the stable ambient-pressure polymorph and its closest experimentally established competitor(s): we consider forms I and II of benzene~\cite{katrusiak2010} and $\alpha$ and $\beta$-succinic acid~\cite{leviel1981} at 100\,K, and $\alpha$, $\beta$, and $\gamma$-glycine~\cite{dawson2005} at 300\,K to compare with available calorimetric data~\cite{perlovich_polymorphism_2001, drebushchak2003}.
The nearly orthorhombic simulation supercells shown in Fig.~\ref{fig:polymorphs}, which contain equivalent numbers of molecules for all polymorphs of the same compound, ensure near-cancellation of centre-of-mass free energies and suffice to converge stabilities with respect to finite-size effects to within 0.1\,kJ/mol (see SI). 

\begin{table}[t]
\begin{center}
 \begin{tabular}{c c c} 
 \hline
 System         & reference data & energy RMSE [kJ/mol] \\
 \hline
 Benzene        & 1,000     & 1.2 \\ 
 Glycine        & 4,000     & 1.6 \\
 Succinic acid  & 2,000     & 2.3 \\
 \hline 
\end{tabular}
\end{center}
\caption{\label{tab:errors}Number of single-point PBE0-MBD calculations underlying each MLP, and their respective root-mean-square errors (RMSE) in predicting energies on a separate test set of configurations from PI simulations of the experimental unit cells.}
\end{table}

\begin{figure}[t]
    \centering
    \includegraphics[width=0.5\textwidth]{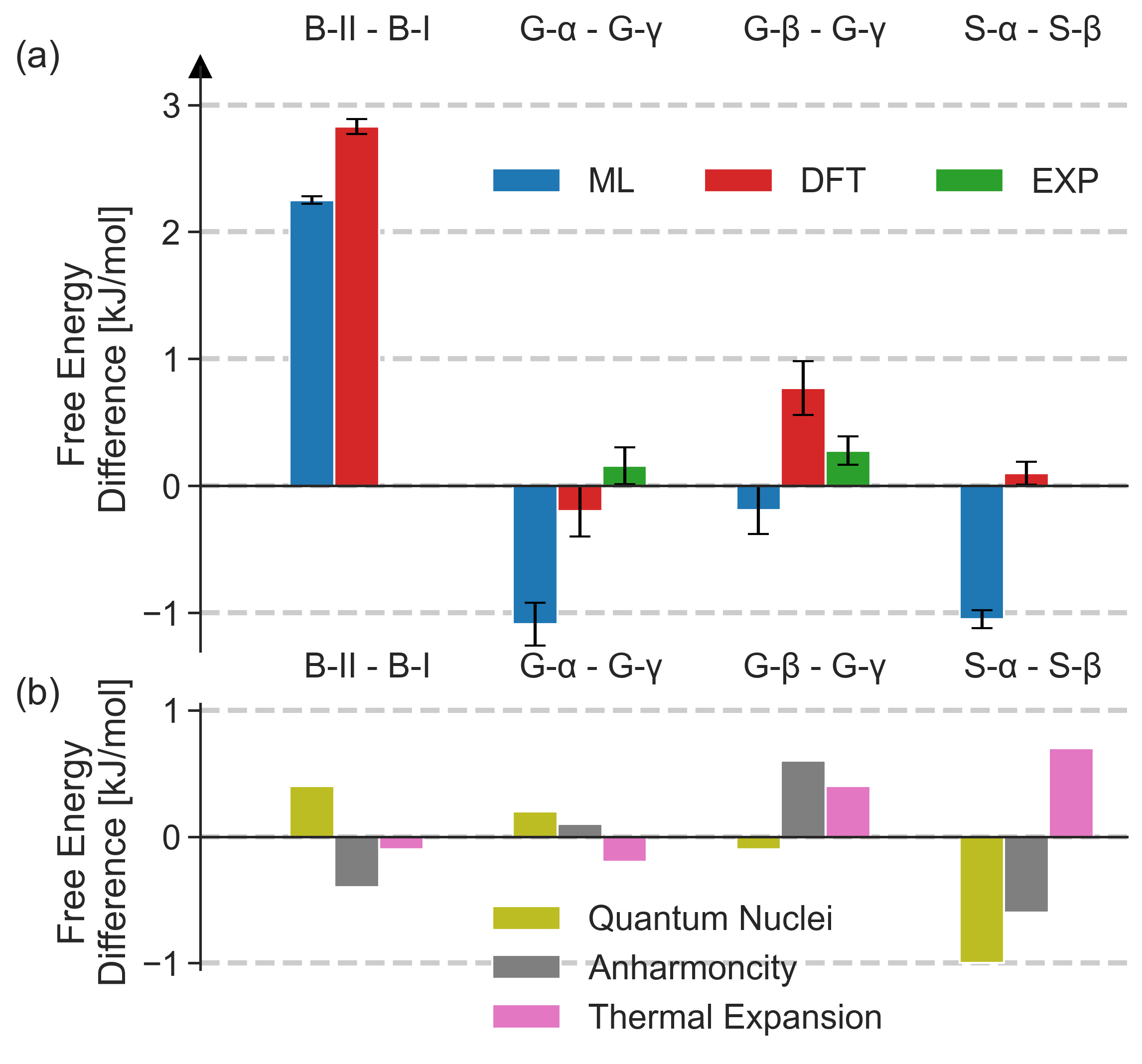}\\
    \caption{Panel (a): Path integral (PI) Gibbs free energy differences between forms II and I of benzene (B-II and B-I), $\alpha$, $\beta$, and $\gamma$-glycine (G-$\alpha$, G-$\beta$ and G-$\gamma$), and $\alpha$ and $\beta$-succinic acid (S-$\alpha$ and S-$\beta$) calculated using PBE0-MBD based machine-learning potentials (MLPs) (blue) with the QTI approach and corrected to the \textit{ab initio} PBE0-MBD DFT level using free energy perturbation (red). Experimental data~\cite{perlovich_polymorphism_2001, drebushchak2003} are shown in green. Panel (b): Contributions of quantum nuclei (olive), anharmonicity (grey), and cell expansion and flexibility (pink) to the relative stabilities of the said polymorphs. These have been respectively obtained by comparing  Gibbs free energy differences to estimates from a classical thermodynamic integration, a harmonic approximation, and a quantum thermodynamic integration using a fixed 0\,K optimized cell. }
    \label{fig:predictions}
\end{figure}

\section*{\textit{Ab initio} thermodynamic stabilities}

As shown in Fig.~\ref{fig:predictions}(a), the final \emph{ab initio} Gibbs free energies (shown in red) reproduce the greater stability of form I over form II of benzene and of $\beta$ over $\alpha$-succinic acid, the metastability of $\beta$-glycine, and the near degeneracy of $\alpha$ and $\gamma$-glycine~\cite{drebushchak2003}. 
Moreover, our Gibbs free energy differences are in agreement with available calorimetry data~\cite{perlovich_polymorphism_2001, drebushchak2003} to within statistical and experimental uncertainties. \\

The QTI approach also yields gradients of Gibbs free energies, including the molar volume, entropy, and heat capacity, which provide indication regarding pressure and temperature-driven changes in relative stability and thus the monotropic or enantiotropic nature of compounds. 
For instance, since molar volumes are derivatives of the free energy with pressure, 
we can predict form II of benzene to become thermodynamically stable over the ambient pressure form I at 1.4\,GPa (at 100\,K), which is in good agreement with the experimentally determined transition pressure of 1.5\,GPa~\cite{yurtseven2013}.
Similarly, we determine the entropy of $\beta$-succinic acid to be smaller than that $\alpha$-succinic acid, making the latter the preferred high-temperature polymorph, in agreement with the \vk{experimental phase behaviour}~\cite{lucaioli_serendipitous_2018}. 
While in the case of glycine, we are only able to predict near degeneracy of $\alpha$- and $\gamma$-glycine at ambient conditions, molar volumes suggest $\alpha$-glycine to be the most stable phase at high pressures, which is in line with experiments showing that it remains stable up to 23\,GPa~\cite{dawson2005}. \\

By comparing rigorous free energies with estimates that exclude nuclear quantum effects (NQEs), anharmonicity, and cell expansion and flexibility, we are able to understand the extent to which these effects and their interplay contributes towards the stability of molecular crystals.
Crucially, as shown in Fig.~\ref{fig:predictions}(b), the size and sign of these effects depends entirely on the compound \emph{and} the polymorphs at hand, highlighting that rigorous QTI is indispensable for predicting phase stabilities and that molecular crystals are typically stabilized by a non-trivial interplay of different physical effects, whose individual importance is belied by the subtle resultant free energy differences.
For instance, the greater stability of form I of benzene hinges on an accurate description the electronic structure, while NQEs and anharmonicity cancel out almost perfectly and thermal expansion affects both forms similarly. 
In contrast, in succinic acid NQEs and anharmonicity cooperatively stabilize the $\alpha$ form and thermal expansion differentiates the two polymorphs.
In glycine NQEs and thermal expansion differently affect the stability of the $\alpha$ and $\beta$ polymorphs with respect to the $\gamma$ form, and neglecting any of the three effects would lead to large errors on the scale of the experimental free energy differences. \\
Meanwhile, the MLP-based stability predictions (shown in blue in Fig. ~\ref{fig:predictions}(a)) are only limited by the accuracy, with which the MLPs reproduce the \emph{ab initio} PES (see Table~\ref{tab:errors}), and consequently correctly reproduce the greater stability of form I of benzene compared to form II. 
At the same time, the incorrect MLP-based stability predictions for succinic acid and glycine highlight the critical importance of the final FEP step. 
Promoting MLP free energies to the \emph{ab initio} level by FEP only incurs the cost of a few tens of \emph{ab initio} energy and force evaluations for configurations sampled by the MLPs.
We note that the cost of this step is comparable to that of common equation-of-state calculations, and thus constitutes a reliable \emph{and} computationally efficient means of predicting the relative stability of polymorphs. \\
Given that errors of 1\,kJ/mol are often considered to be within ``chemical accuracy'', it is worth emphasising that the compounds considered here are not hand-picked, ``pathological'' examples, but expected to be representative of many bio-molecular compounds. 
The small free energy differences between polymorphs, which are smaller than $k_B T$ but can be resolved experimentally~\cite{perlovich_polymorphism_2001, drebushchak2003} thanks to the kinetic suppression of interconversion between polymorphs, constitute a very stringent test of our framework and its ability to accurately capture phase stability. 
By matching the sub-kJ/mol accuracy of calorimetry experiments, it provides a robust foundation for studying transition temperatures, pressures, and rates, and permits benchmarking sophisticated electronic structure theories against experiment. 

\section*{Comparison with approximate approaches}

In order to further highlight the advantages of the approach proposed here over established 
approximate methods for ranking stabilities in CSP, we assess the limitations of the most widely used approximate methods, prefaced by acknowledging their successes for a wide range of applications~\cite{hoja_reliable_2019}. 
We note that the MLPs reproduce the \textit{ab initio} PESs with sufficient accuracy to assess the impact of approximations to nuclear motion, and compare the respective approximate (free) energy differences between polymorphs to the corresponding exact MLP Gibbs free energies. \\

\begin{figure}[t]
    \centering
    \includegraphics[width=0.5\textwidth]{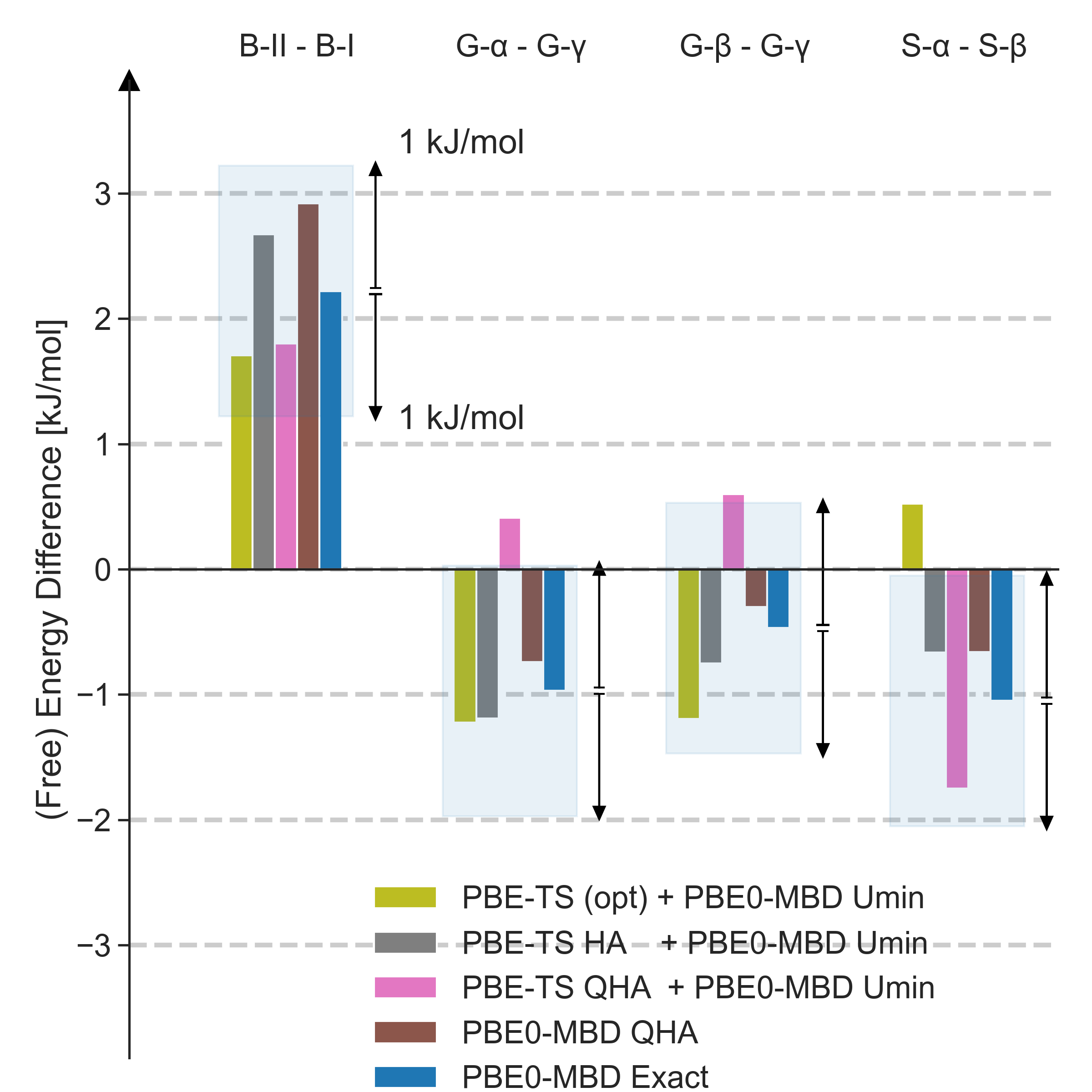}\\
    \caption{MLP (free) energy differences between forms II and I of benzene (B-II and B-I), $\alpha$, $\beta$, and $\gamma$-glycine (G-$\alpha$, G-$\beta$ and G-$\gamma$), and $\alpha$ and $\beta$-succinic acid (S-$\alpha$ and S-$\beta$) at different tiers of accuracy: fixed-cell optimization using PBE-TS with a PBE0-MBD single point energy (green), fixed-cell optimization and harmonic free energy using PBE-TS   with a PBE0-MBD single point energy (gray), quasi-harmonic approximation (QHA) free energy using PBE-TS with a single-point PBE0-MBD correction (pink), full PBE0-MBD based QHA (brown), and the exact PI free energy difference (blue). The shaded region indicates free energy differences within 1\,kJ/mol of the respective exact PI result as a guide to the eye.
    \label{fig:tiers}}
\end{figure}

The current state-of-the-art is to correct (free) energies 
on the basis of a single-point hybrid-functional DFT calculation for the structure relaxed using semi-local DFT~\cite{lucaioli_serendipitous_2018}.
Thermal and quantum nuclear effects are included within a harmonic approximation (HA)~\cite{reilly_role_2014}, while thermal expansion is modeled by relaxing the cell within a quasi-harmonic approximation (QHA)~\cite{hoja_reliable_2019}.
These corrections are generally computed at the semi-local DFT level. 
As shown in Fig.~\ref{fig:tiers}, these approaches neither universally predict the most-stable form (as they exhibit errors larger than 1\,kJ/mol), nor systematically converge to the full hybrid-functional QHA reference. 
This highlights the need to go beyond a single-point hybrid-functional DFT correction to semi-local configurational or (quasi-)harmonic free energies to consistently deliver correct stability orders and free energies differences with sub-kJ/mol accuracy.
We further note that the above results benefit substantially from the fortuitous cancellation of errors~\cite{kapil_assessment_2019}, but the residual errors cannot be estimated and \emph{apparent} physical insights may be misleading. \\ 
Since the hybrid-functional-based QHA seems to be competitive with the rigorous PI approach, is further worthwhile to put the cost of the calculations into perspective. 
For glycine, as the most costly example, the 4,000 PBE0-MBD calculations on \vk{\emph{unit cells}} constituting the reference data for the MLP, the MLP-based PI thermodynamic integration, and the 50 PBE0-MBD calculations on \vk{\emph{supercells}} required for the FEP contribute roughly equally to the total cost of around 148,000 core-hours per polymorph.
For comparison, computing PBE0-MBD HA free energies for the same simulation supercells using finite differences and non-diagonal supercells to probe individual $k$-points~\cite{lloyd_2015}, but not leveraging the MLP, would require about thrice the core-hours. 
A PBE0-MBD QHA free energy calculation would be an order of magnitude higher in computational cost. 
Although (Q)HA free energies may also be computed inexpensively using MLPs, they \emph{cannot} be promoted to their first-principles counterparts in a straight-forward and cost-effective manner as exact MLP free energies.
Despite a focus on universal applicability over efficiency, the cost of the above rigorous Gibbs free energies is thus small compared to the estimated cost of calculating free energies within the (Q)HA using hybrid-functional DFT. 

\section*{Discussion}

%
The ability of our approach to predict free energy differences with sub-kJ/mol accuracy, renders it valuable in identifying ``competing'' polymorphs with similar lifetimes to the most stable form.
It bridges the gap between theory and experiments by allowing direct comparison of free energy differences with calorimetric data -- a significant improvement over current approaches, which require error-prone \textit{ad hoc} extrapolations to 0\,K~\cite{zen_fast_2018}.
Moreover, (i) rigorous predictions of the entropy, molar volume, and heat capacity and (ii) robust MLPs with \emph{ab initio} accuracy come as complements of our approach.
The former are directly related to ``thermodynamic rules of thumb'', which are widely used by experimentalists to assess stability trends~\cite{lee2014}, while the latter enable structure determinations for experimental samples based on 
NMR~\cite{engel_importance_2021-1} and vibrational spectra~\cite{shepherd_efficient_2021}.
Furthermore, a rigorous account of thermally-induced phase-transitions can be obtained without repeating the procedure at every state point. 
The combination of QTI with parallel tempering~\cite{sugita_replica-exchange_1999}, can enhance the efficiency of performing single temperature (or pressure) sweeps, yielding full phase diagrams, while also sampling ``slow'' degrees of freedom such as conformational transitions, inaccessible to approximate methods~\cite{kapil_assessment_2019}.
All these features are highly sought-after by the pharmaceutical industry, as they are made possible at a manageable computational cost. \\
Our protocol easily extends to stability predictions for other complex molecular crystals, 
as its data-driven nature accelerates MLP development, irrespective of the material under consideration.
As a proof of this, we have developed MLPs for polymorphs of three complex pharmaceuticals -- aspirin, paracetamol, and XXIII, the most complicated system~\cite{hoja_reliable_2019} from the latest blind test of organic crystal structure prediction methods~\cite{reilly_report_2016} -- and tested them by performing PI simulations in the constant pressure ensemble, as required for QTI (see SI). 
Although these MLPs have been trained on DFTB data as a proof of concept and consequently lack chemical accuracy, they remain robust and capture the molecular flexibility of these systems. 
Given that dynamic disorder, thermal expansion, conformational relaxation of the molecular units, and potential (dynamic) instabilities of candidate polymorphs are automatically accounted for within the QTI approach, we expect stability predictions to be very robust with respect to the nature of the candidate polymorphs, and thus directly applicable to said pharmaceutical and blind-test systems. \\ 

In applications involving large numbers of polymorphs or polymorphs with large unit cells, suitable sets of reference configurations
can be generated based on configurations of liquid or amorphous states at different pressures~\cite{monserrat2020}. 
This exploits that the accuracy of MLPs, which predict energies and forces on the basis of local contributions, rests on having reference data for all distinct local atomic environments~\cite{monserrat2020}, rather than for all polymorphs of interest.
The computational cost of building the training set then remains largely independent of number and unit cell size of the polymorphs of interest. 
For large numbers of polymorphs the cost per polymorph thus effectively reduces to that of the MLP-based thermodynamic integration and of FEP.
In practice, the computational cost of FEP can be reduced by only running as many \emph{ab initio} calculations, as required to reduce the statistical error to below the predicted free energy differences between polymorphs. 
For instance, less than a handful of PBE0-MBD calculations would have sufficed to conclusively establish that form I of benzene is more stable than form II. 
Indeed, subject to estimates of the uncertainty of the MLP predictions~\cite{musil_2019, imbalzano_uncertainty_2021}, it may be possible to omit FEP altogether. 
Recent work on the use of higher body-order correlations in atomistic representations~\cite{nigam_recursive_2020}, and on including long-ranged interactions~\cite{grisafi_incorporating_2019} promises to enable sub-kJ/mol accuracy, eliminating the need for FEP even in applications involving subtle free energy differences. \\

Finally, the empiricism involved in selecting the exchange-correlation functional and dispersion correction used in the DFT calculations, can be removed by using PESs evaluated using beyond-DFT electronic structure theory. 
Crucially, our scheme extends naturally to predictions of Gibbs free energies based on quantum-chemical electronic structure methods~\cite{cervinka_ab_2018, zen_fast_2018} such as MP2, RPA, coupled cluster or quantum Monte Carlo, some of which are systematically improvable, and can thereby be rendered truly \emph{ab initio}~\cite{chmiela2018}. 
While these come at an increased computational cost per calculation, 
recent developments in machine learning for materials science~\cite{schran_committee_2020}, promise to minimize the number of quantum-chemical calculations required to train accurate MLPs and thus to keep the overall costs in check. 
Indeed, recent work demonstrates the corresponding construction of robust and accurate MLPs for CCSD(T) reference data~\cite{chmiela2018}. \\


In conclusion, marrying state-of-the-art electronic structure, free energy, and machine-learning methods in a widely applicable framework enables rigorous, predictive free energy calculations for complex (organic) molecular crystals at general thermodynamic conditions. 
The unprecedented accuracy of our approach sets the stage for future studies of kinetic effects as well as full $p$-$T$ phase-diagrams in a reliable and computationally efficient manner, paving the way for guiding experimental synthesis of such materials. 
The protocol and the scripts provided as supplementary material permit its application practically out-of-the-box.
Determining the relative stability of generic polymorphic compounds is a recurrent problem across different domains of science and engineering -- from nucleation theory to the practical design of pharmaceuticals -- and we hope that the robust and easy-to-use nature of our end-to-end protocol will facilitate reliable, accurate free energy calculations beyond the computational chemistry community. 
%

\section*{Methods}
\label{sec:methods}

{\small

\noindent {\bf Machine-learning potentials:}
We have constructed Behler-Parinello type neural network 
potentials~\cite{behl-parr07prl} for benzene, glycine, and 
succinic acid using the n2p2 code~\cite{singraber_library-based_2019}.
In this framework, structures are encoded in terms of local 
atom-centered symmetry functions (SF)~\cite{behl-parr07prl}. 
Initial sets of SFs were generated following the recipe of
Ref.~\cite{imbalzano2018cur}. Based on the same reference
structure-property data subsequently used for training,
the 128 (benzene and succinic acid) and 256 (glycine) most
informative SFs were extracted 
via PCovCUR selection~\cite{cersonsky2021}. \\

\noindent Our data is based on Langevin-thermostatted PI NVT simulations 
at 300\,K, performed using the i-Pi force engine~\cite{kapil_2018_ipi} 
coupled to DFTB+~\cite{noauthor_dftb_nodate} 
calculations with the 3ob parametrization~\cite{kruger_validation_2005}.
For each polymorph multiple cells were simulated, rescaling 
the experimental cell lengths and angles by up to 10\% and 5\%, 
respectively. 
\vk{The trajectories of PI replicas for all polymorphs} of a given compound were concatenated and 
farthest-point sampled~\cite{eldar_1997_fps, ceriotti_2013_sketchmap, campello_2015_fps} to extract the most 
distinct configurations for feature selection and MLP training.
Subsequently, \emph{ab initio} reference energies and forces 
were evaluated for said configurations. \\

\noindent To minimise the computational cost of the reference calculations 
the MLPs are composed of a baseline potential trained to 
reproduce energies and forces from more affordable PBE-DFT~\cite{perdew1996pbe} 
calculations with a TS dispersion correction~\cite{tkatchenko2009ts} 
(PBE-TS), and a $\Delta$-learning~\cite{ramakrishnan_big_2015} correction trained 
(on ten times fewer training data) to reproduce the difference 
between the baseline and more expensive calculations with the 
hybrid PBE0 functional~\cite{perdew1996pbe0, adamo1999pbe0} 
and the MBD dispersion correction~\cite{tkatchenko2012mbd, ambrosetti2014mbd} (PBE0-MBD). 
For a separate test set, the MLPs reproduce PBE0-MBD energies 
with root-mean-square errors of 1.2\,kJ/mol for benzene, 
1.6\,kJ/mol for glycine, and 2.3\,kJ/mol for succinic acid, 
respectively. \\

\noindent {\bf {\emph Ab initio} DFT calculations:} 
\vk{PBE0+MBD} calculations were performed using FHI-aims~\cite{blum2009aims, ren2012aims, levchenko2015aims} with the standard FHI-aims 
``intermediate'' basis sets and a Monkhorst-Pack k-point grid~\cite{monkhorst1976} with a maximum spacing of 
$0.06 \times 2\pi \Ang^{-1}$.
The PBE-TS baseline calculations for a $\Delta$-learning approach 
were performed using Quantum Espresso v6.3, the same k-point grid, 
a wavefunction cut-off energy of 100\,Rydberg, and the optimised, 
norm-conserving Vanderbilt pseudopotentials from Ref.~\cite{schlipf2015}.  \\

\noindent {\bf Free energy methods:}
For each polymorph the average cell was determined using MLP based 
path integral (PI) NST simulations~\cite{raiteri_reactive_2011} at the desired inverse temperature $\beta$,
accounting for anharmonic quantum nuclear motion and anisotropic cell
fluctuations. The difference between the Gibbs and Helmholtz free energies
is computed from a MLP based PI NPT simulation based on its average cell is
\begin{equation}
    \gml(P^{\text{ext}}, \beta) - \aml(V,\beta) = P^{\text{ext}} V + \beta^{-1} \ln\rho(V | P^{\text{ext}}, \beta),
    \label{eq:g}
\end{equation}
where $\rho(V | P^{\text{ext}},\beta)$ is the probability of observing the cell 
volume $V$ at external pressure $P^{\text{ext}}$ and inverse temperature $\beta$.
A standard Kirkwood construction~\cite{frenkel_new_1984} that transforms the Hamiltonian from a 
harmonic to an anharmonic one provides the  difference between the 
anharmonic and the harmonic quantum Helmholtz free energies:
\begin{equation}
    \aml(V,\beta) - \amlhar(V,\beta) = \int_0^1 d\lambda \avg{\hat{H}_{\text{MLP}} - \hat{H}^{\text{har}}_{\text{MLP}}}_{V,\beta, \hat{H}_\lambda},
    \label{eq:aanhar}
\end{equation}
where $\hat{H}_\lambda$ is the Hamiltonian of the MLP 
alchemical system with the potential $U_\lambda \equiv \lambda \uml + (1-\lambda) \umlhar$, 
and $\avg{\cdot}$ is the ensemble average computed from a PI NVT simulation.
The reference absolute harmonic Helmholtz free energy is 
obtained from a harmonic approximation using
\begin{equation}
    \amlhar(V,\beta) = \uml(V) + \sum_{i} \left[ 
    \frac{1}{2} \hbar \omega_i + \beta^{-1} \ln\left( 1 - e^{- \beta \hbar \omega_i} \right)
    \label{eq:ahar}
    \right],
\end{equation}
where $\omega_i$ is the frequency of the $i$-th phonon mode. 
In a final step, the \emph{ab initio} Gibbs free energy is 
obtained from its MLP counterpart by free energy perturbation using
\begin{equation*}
        \gdft(P^{\text{ext}},\beta) - \gml(P^{\text{ext}},\beta) 
        = - \beta^{-1}\ln\avg{ e^{-\beta ({\udft-\uml})}}_{P^{\text{ext}},\beta,\hat{H}_{\text{MLP}}}.
    \label{eq:fep}
\end{equation*}
For systems exhibiting large amplitude curvilinear motion, the harmonic-to-anharmonic thermodynamic integration can be performed efficiently using a Pad\'e interpolation formula~\cite{rossi_anharmonic_2016}. \\

\noindent {\bf Understanding the role of different effects:}
\vk{We disentangle the role of anharmonicity directly from Eq.~\ref{eq:aanhar}, and that of thermal expansion by comparing the Helmholtz free energies from Eq.~\ref{eq:aanhar} for the variable-cell geometry-optimized and mean PI NST cells. The role of the quantum nature of nuclei is quantified by comparing the classical and quantum Gibbs free energies. We calculate the former using the Helmholtz free energy of the classical harmonic oscillator as a reference, and evaluating Eqs.~\ref{eq:g} and \ref{eq:aanhar} using classical molecular dynamics.} \\

\noindent {\bf Free energy gradients:}
\vk{Volume and entropy are related to gradients of the free energy 
\begin{equation}
    V = \left(\frac{\partial G}{\partial P}\right)_{N,T},\quad S = -\left(\frac{\partial G}{\partial T}\right)_{N,P} .
\end{equation}
Differences between equilibrium (molar) volumes of polymorphs can directly be observed in PI NPT simulations. Meanwhile entropic differences can be computed from 
\begin{equation}
    S = -\frac{1}{T} (G - H),
\end{equation}
with $G$ from Eq.~\ref{eq:g} and the enthalpy $H$ from the associated PI NPT simulation.
Linear extrapolation then permits estimating if and at which pressures $P_c$ and temperatures $T_c$ the Gibbs free energy difference between polymorphs will vanish and a phase transition should be expected:
\begin{equation}
    \Delta G = -(T_c - T) \Delta S + (P_c - P) \Delta V
\end{equation}}
}

\section*{Acknowledgements}
\label{sec:acknowledgements}

\noindent VK acknowledges funding from the Swiss National Science Foundation (SNSF), Project $\text{P2ELP2}\_\text{191678}$, and support from the NCCR MARVEL (funded by the SNSF) and Churchill College, University of Cambridge.
EAE acknowledges funding from Trinity College, Cambridge.
The authors thank Benjamin Shi, Daan Frenkel, Angelos Michaelides, Sally Price and Michele Ceriotti for valuable suggestions on the manuscript.
EAE and VK acknowledges allocation of CPU hours by CSCS under Project IDs s960 and s1000.

\end{document}